\documentclass{article}
\usepackage{graphicx} % Required for inserting images
\usepackage{changepage}
\usepackage{moreverb}
\usepackage[hyphens]{url}
\usepackage{multirow}
\usepackage{ulem}
\usepackage[colorlinks,bookmarksopen,bookmarksnumbered,citecolor=red,urlcolor=red]{hyperref}

\title{Artificial Intelligence across Europe: A Study on Awareness, Attitude and Trust}
\author{Teresa Scantamburlo, Atia Cortés, Francesca Foffano, Cristian Barrué,\\ Veronica Distefano, Long Pham, Alessandro Fabris}

\date{15 May 2023}

\begin{document}

%\keywords{Trustworthy AI, Europe, Society, Public Perception, AI Policy}

\maketitle

\begin{abstract}
This paper presents the results of an extensive study investigating the opinions on Artificial Intelligence (AI) of a sample of 4,006 European citizens from eight distinct countries (France, Germany, Italy, Netherlands, Poland, Romania, Spain, and Sweden). The aim of the study is to gain a better understanding of people’s views and perceptions within the European context, which is already marked by important policy actions and regulatory processes.
To survey the perceptions of the citizens of Europe we design and validate a new questionnaire (PAICE) structured around three dimensions: people's awareness, attitude, and trust. We observe that while awareness is characterized by a low level of self-assessed competency, the attitude toward AI is very positive for more than half of the population. Reflecting upon the collected results, we highlight implicit contradictions and identify trends that may interfere with the creation of an ecosystem of trust and the development of inclusive AI policies. The introduction of rules that ensure legal and ethical standards, along with the activity of high-level educational entities, and the promotion of AI literacy are identified as key factors in supporting a trustworthy AI ecosystem. We make some recommendations for AI governance focused on the European context and conclude with suggestions for future work.
\end{abstract}

\section{Introduction}
In April 2021 the European Commission (EC) proposed a set of rules to regulate Artificial Intelligence (AI) systems operating across Europe \cite{ai_act}. This was an important step in a long-term process in which the European Union developed its approach towards AI \cite{krarup2023european}, setting up policy agendas \cite{ai_agenda} and ethics guidelines \cite{Trustworthy_EU} among others. During this period, the EC sought feedback from different stakeholders to ensure an inclusive policy development, such as the consultation run from February to June 2020 to gather opinions about the White Paper on AI \cite{Whitepaper_EU}. Usually, these consultations solicit reflection on specific actions or policy proposals and can reveal partial information, if anything, about what people think about AI and its related impact on society. 

Knowing people's views and perceptions is key to deploying effective governance mechanisms and integrating rules into society. In this paper, we aim to fill this gap and report the results of a survey investigating the knowledge and perception of AI in Europe. For this reason, we designed, developed, and validated a new questionnaire, the Perceptions on AI by the Citizens of Europe questionnaire (PAICE), structured around three dimensions: Awareness, attitude, and trust. Based on a computer-assisted web interview methodology (CAWI) we collected and analyzed the opinions of 4,006 European citizens from eight countries (France, Germany, Italy, Netherlands, Poland, Romania, Spain, and Sweden), stratified by age, gender, and geographic regions.

The collected responses show that respondents' self-assessed knowledge about AI is low, while their attitude is very positive and slightly varies depending on the context of use (e.g. approval is lower when AI is applied to human resources management). The most important measures to increase trust in the AI ecosystem include the introduction of laws by national authorities, transparent communication by AI providers, and education activities. Among trusted entities that could ensure a beneficial use of AI, universities and research centers are ranked higher than other organizations (e.g. national governments and tech companies). The statistical analysis shows that the questionnaire has good internal consistency and that the validity is adequate.

We analyze the results of the survey and identify a few contrasting perceptions which may reflect three broader social trends: 1) approval of a hyped, but poorly known, technology; 2) disconnect from public AI policies; 3) poor engagement with AI education and training. We discuss how these trends may create friction in the creation of a Trustworthy AI culture and suggest a few recommendations. Our findings call for greater consideration of people's views and participation in AI policy-making, especially if we consider the rapid transformations introduced by AI into society and the abundance of policy efforts by states and intergovernmental organizations \cite{oecdAI, ai_act, cheng2022shaping, AlgoAct, AfricaAI, simbeck2022facct}. 

%(European Perception on AI - EPAI) \textcolor{red}{European Citizens' Perception on AI Survey - ECPAIS // Survey on the Perception of AI by the Citizens of Europe - SPAICE} 
%(for more details see section \ref{sec:reserch_quest}. 

\subsection{Related work}

AI and trust recall a vast academic literature investigating shared principles among ethics guidelines \cite{hagendorff2020ethics,gille2020we,hongladarom23}, as well as challenges and future directions \cite{siau2018building,lockey2021review,jacovi2021formalizing,glikson2020human,lee2018understanding}. In this section, we focus on previous surveys analyzing citizen awareness, trust, and attitude towards AI from different perspectives. 

In a global study surveying 10,000 citizens spanning eight countries across six continents \cite{kelley2021exciting}, respondents reported a mix of positive and negative feelings about AI. In a similar study, the UK expressed a markedly negative view of AI, while showing a reasonable understanding and awareness of this technology \cite{cave2019scary}. The US population has been surveyed on a key dimension of trust: the perception of governance \cite{zhang2019artificial,zhang2020us}. While most people (especially older segments) find the issue very important, they state that they have little trust in the actors who have the power to develop and manage AI (e.g. companies, universities, US agencies). Another US-related work investigated the ethical preferences of different groups of people and found that AI practitioners' value priorities differ from those of the general public \cite{jakesch2022different}. 

Studies focused on the perception of AI in Europe are not entirely new. In an EU-wide survey, the authors focused on a notion of AI centered around robotics, finding attitudes to be generally positive, with concerns related to job losses \cite{european2017special}, later confirmed in a follow-up study \cite{european2021special}. These are generic studies of EU public opinion about science and technology, with only a marginal focus on AI. %Our survey focuses on a broader definition of AI.
A subsequent survey on opinion about AI highlighted discrimination and lack of accountability as key concerns for European citizens, and a belief that public policy intervention is needed, shared by a majority of respondents \cite{european2019}.

Recently, \cite{kerr2020expectations} analyzed the positive and negative expectations of 164 individuals visiting a Science Gallery exhibition in Dublin. The study found that awareness of AI is relatively good, opportunities are related to economic growth and social progress (e.g. mentioning the positive impact on medicine, science, and environments) and concerns are connected to automation, followed by privacy and surveillance. \cite{sartori2022} examined awareness of AI, emotional responses to narratives, and the perceived likelihood of future scenarios in Italy. The authors pointed out a positive correlation between the level of digital expertise and general knowledge of AI and showed an important gender divide with respect to the emotional response to narratives with women more concerned than men across all scenarios. \cite{kieslich2022artificial} investigated how German people prioritize different ethical principles (transparency, fairness, non-maleficence, responsibility, beneficence, privacy, and machine autonomy) with regard to the application of AI to fraud detection. The study found that all ethical principles are equally important for the respondents but different preference profiles for ethically designed systems exist.

\subsection{Research questions}

The present work departs from the existing literature in two fundamental ways. First, it takes AI as its main target not as part of broader investigations in science and technology \cite{european2021special}, connecting different perspectives (such as awareness and trust) with specific use cases. Second, it aims at reaching a large population involving more than one European country or demographics \cite{kieslich2022artificial, sartori2022, kerr2020expectations}. 

The questionnaire was developed in the context of %the European project AI4EU 
an Horizon 2020 project by a multidisciplinary team of researchers. The research questions addressed by the team are the following:

\begin{itemize}
    \item \textbf{RQ1}: \emph{To what extent are EU citizens familiar with AI and the surrounding debate?} This covers aspects concerning citizens' awareness and competency such as: what people think they know about AI, where they think AI is applied, what is the perceived impact of AI, and which EU initiatives addressing ethical and legal concerns they are aware of.
    \item \textbf{RQ2}: \emph{To what extent do EU citizens approve AI?} This research question connects to citizens' attitude towards AI and its use in some specific sectors or contexts of application (such as job recruitment). 
    \item \textbf{RQ3}: \emph{What could contribute to increasing citizens' trust?} This question investigates citizens' priorities to promote the responsible development of AI in terms of actions, actors and ethical requirements.
\end{itemize}
These questions guided the development of the questionnaire around the dimensions of awareness, attitude, and trust. The structure of the questionnaire was also explored in our analysis (i.e. validity and reliability). This allowed us to identify which items of the questionnaire can be used to validate the dimensions suggested by the team of experts who designed the research instrument.

The rest of the paper is organized as follows. First, we present the methodology guiding the survey design. Next, we report the results of the survey according to the dimensions of interest (i.e. awareness, attitude, and trust), and analyze the validity and reliability of the questionnaire. We discuss the results pointing out implicit tensions and discussing potential barriers to the development of inclusive AI policy processes, thereby making recommendations to improve current efforts, especially at the European level. Finally, we summarise our findings and suggest future research directions.  

\section*{Methods and materials}

\subsection*{Survey method}% Teresa atia e veronica 

This survey was conducted by the market research agency Marketing Problem Solving (MPS) based in Italy \cite{Mpswebsite}. The survey was carried out through online interviews (CAWI) on the basis of a structured questionnaire. The average completion time was 20 minutes. MPS programmed the script of the questionnaire through the creation of a website hosted on the web server owned by MPS and managed the data collection process. 

The invitation to fill out the questionnaire was sent by e-mail to members of an online panel who voluntarily agreed to share their opinions. To facilitate the task, panel members received the questionnaire in their own language. The respondents were free to drop out at any point and had the opportunity to go back to previous items and change their answers. Respondents' information was recorded in compliance with the General Data Protection Regulation (GDPR) and the Italian legislation on data protection and privacy. 

From the 1st to the 15th of June 2021 MPS, realized a total of 4,006 interviews in eight European countries: France, Germany, Italy, Netherlands, Poland, Romania, Spain, and Sweden. Countries were selected with a view to cover different European regions (southern, central/eastern, northern, and western). The survey was completed by individuals aged between 18 and 75 years. Quotas were imposed to ensure the representativeness of the sample with respect to gender, age group (18-34, 35-54, 55-75), and geographical area of residence.

Before undertaking the survey, MPS tested the questionnaire with a sample of panel members to assess the clarity of instructions and the average completion time. MPS monitored the whole interview process to ensure the quality of responses, e.g. by removing participants who completed the survey too quickly or provided contradictory answers.
%For example, completion time less than half of the average time is one of the criteria used by MPS to assess the quality of interview. 

The original version of the survey was developed and revised in English and then translations in other languages (Italian, Spanish, German, Polish, French, Romanian, Dutch, Swedish) were made by professional translators or mother tongue experts.  
%MPS Continued to recruit respondents until reaching the target of 500 interviews per country.

\begin{table*}
\centering
  \caption{Description of the survey population}
  \label{table1}
%   \begin{tabular}{l+r|ccc|cccc}
% \hline
% \multicolumn{ 1}{l|}{ } & \multicolumn{ 2}{|c}{Gender} & \multicolumn{ 3}{|c}{Age groups}  & \multicolumn{ 3}{|c}{City size by population}\\
% \hline
% \multicolumn{ 1}{l|}{ } & {woman & man} & {18-34} & {35-54} & {55-75 & $<$ 10K & 10K-100K & $>$ 100K} \\
% \hline
\begin{tabular}{l|cc|ccc|ccc}
\hline
    \multirow{2}{*}{Country} &
    \multicolumn{ 2}{|c}{Gender} & \multicolumn{ 3}{|c}{Age groups}  & \multicolumn{ 3}{|c}{City size by population} \\
    & woman & man & 18-34 & 35-54 & 55-75 & $<$ 10K & 10K-100K & $>$ 100K \\
    \hline

$France$ & 48.5\% & 51.5\% & 28.9\% & 38.4\% & 32.7\% & 42.2\% & 33.9\% & 23.9\%\\

$Germany$ & 50.1\% & 49.9\% & 27.9\% & 38.3\% & 33.7\% & 30.2\% & 36.7\% & 33.1\%\\

$Italy$ & 48.2\% & 51.8\% & 23.8\% & 41.0\% & 35.2\% & 22.8\% & 51\% & 26.2\%\\ 

$Netherlands$ & 50.2\% & 49.8\% & 28.6\% & 36.6\% & 34.8\% & 19\% & 49\% & 32\%\\

$Poland$ & 49.0\% & 51.0\% & 30.8\% & 37.4\% & 31.8\% & 24.2\% & 31.6\% & 44.2\%\\

$Romania$ & 50.0\% & 50.0\% & 27.5\% & 40.9\% & 31.5\% & 29.3\% & 28.2\% & 42.5\%\\

$Spain$ & 49.7\% & 50.3\% & 21.6\% & 44.5\% & 33.9\% & 19\% & 41.1\% & 39.9\%\\

$Sweden$ & 50.0\% & 50.0\% & 31.9\% & 37.9\% & 30.1\% & 12.4\% & 45.5\% & 42.1\%\\

\hline
\end{tabular}
\end{table*}

\subsection{Population}%veronica e teresa
 %\new{[Alessandro: Seems somewhat strange to describe the data with past tense, may want to consider switching to present]}

To obtain a random stratified sample, members of the population were first divided into non-overlapping subgroups of units called \emph{strata} (country), then, a sample was selected from each stratum based on geographic regions (unit of analysis), age groups and gender through a simple random sampling. The sample was made up of 4,006 individuals with equal representation for each country (12.5\%). 
%\remove{MPS kept recruiting individuals until a representative sample was reached.}

The sample was composed of individuals in the age range 18-75 (mean age = 45, std = $\pm{14.83}$) where women were $49.3\%$ (mean age = 46, std = $\pm{14.92}$) and men $50.4\%$ (mean age = 45.5, std = $\pm{14.83}$). Note that in our analysis we considered only male and female groups since the respondents choosing the option ``others'' were only 0.3\% of the whole population. The age groups were coded into three levels: young (18-34 years), middle age (35-54), and senior (55-75) people. In particular, 26.5\% of respondents were young, 39.4\% were middle age, and the remaining 33\% were senior. We also investigated the population size of the place of residence and found that 25\% of respondents lived in a city with a population up to 10K, about 40\% lived in a city with 10-100K inhabitants, and the remaining 35\% lived in a large city with more than 100K inhabitants. Information about gender, age groups, and city size related to respondents of each country is summarized in Table \ref{table1}.

With reference to formal education, the descriptive analysis highlighted that 40\% of the respondents had the highest level of formal education (bachelor, master, or doctoral degree). Note that this percentage is higher than the share of European citizens with tertiary education (i.e. also including trade schools and vocational education) which is estimated at 31\% \cite{eurostat2021educational}. The choice of the survey methodology, based on online interviews, possibly facilitated the participation of subjects with higher levels of education.  
%Regarding the job, the greatest part of the population is formed by employee or retired people (28.7\% and 17.4\% respectability). 
%\textcolor{blue}{Similarly to above, either compare this sample value with the respective population value, or avoid mentioning.} 

To investigate confidence with Information and Communication Technology domains we submitted to the respondents an item assessing their level of competence in digital skills on a five-point ordinal scale from almost no knowledge to advanced knowledge. It was observed that 44\% of the respondents have an intermediate level of competence in digital skills. Among those who feel less competent in digital skills, we found French and German respondents who represent respectively 31.7\% and 34.5\% of the population surveyed in each country. The countries reporting the highest level of competency are Spain and Italy where respondents with intermediate or advanced knowledge are 82.9\% and 79.8\% respectively. For more details on digital skills and formal education see tables ``digital skills'' and ``education'' in the supplemental material (Digital skills).

%Moreover, it is noteworthy that the respondents of France and Germany who said they did not have or possess a basic level of digital skills are respectively 31.7\% and 34.5\% of the population surveyed in each. On the other hand, the respondents in Spain and Italy showed  the highest level of competence (from intermediate to advanced knowledge) in digital skills with 82.9\% and 79.8\% . In terms of population size, 56.8\% of respondents live in the  city, which have a population of at least 50.000 inhabitants. %In particular 853 (21.3\%) belonged to the city with a population in the 30000 to 100000 range of people and 826 (20.6\%) belonged the city with a population of over 250000 inhabitants. 
%The rest 43.2\% respondents belonged  to the sub-urban areas with a population that does not exceed 30000 inhabitants.}

\subsection{Questionnaire design}
The PAICE questionnaire was created
%  within the  European project AI4EU project \texttt{redacted} 
by a group of researchers from different backgrounds (AI \& Computer Science, Philosophy, Engineering, Psychology, and Communication) including the authors of the present work. 
%recruited through the project network and social media channels.
%sharing a common interest in AI ethics and European AI policy. %They were recruited through the AI4EU network and a public call promoted on the project website and social media channels.
%\cite{AI4EUworking}

The design of the questionnaire took six months, from January to June 2021, during which the group met on a monthly basis. In the early stages of the design process, the group collected and analyzed the existing literature and previous surveys at a European and worldwide level. Based on the literature review, the group identified the research questions and subsequently defined the questions for the research instrument. 
After a refinement process, the group agreed on a total of 14 items including Likert scale, dichotomous, multi-response items, and ranking. The items were organized according to the three dimensions introduced above (awareness, attitude, and trust) with a view to address the starting research questions. An overview of the structure of the questionnaire with question types and the topics of each item is reported in Table \ref{tab:paice}. 
Note that some questions, since they applied to different sectors, policy measures, or entities, were split into sub-items (e.g. Q7\_1 to Q7\_10). 

In addition, the questionnaire presented: a control question about the perceived impact of AI, a question investigating the interest in attending a free course on AI, and seven questions on socio-demographic aspects (i.e. age group membership, gender, geographical area, population size, job sector, level of education, and digital expertise). The control question, which was a repetition of item Q3 (see Table \ref{tab:paice}), was added to assess possible changes in opinions after the completion of the questionnaire. The English version of the full questionnaire is available in the supplemental material (Questionnaire). 
%46 items comprising: 39 Likert scale items, 1 repeated Likert scale item (control questions), 4 dichotomous items, 1 multiple-response item and 1 ranking.  

Likert scale items ranged from 1 to 5 where 1 referred to negative or low values (e.g. ``not at all'', ``never'', ``not important at all'' and ``strongly disapprove'') and 5 to positive or high values (e.g. ``a lot'', ``always'', ``very important'' and ``strongly approve''). For item Q5 we also added the option ``I don't know'' to accommodate respondents who did not have a clear opinion on the topic (awareness of interaction). 
We chose the 5-point Likert scale because this is largely used in social science research to study human attitudes and perceptions \cite{nunnally1994}. Though the optimum number of choices in a Likert-type scale is a subject of dispute \cite{joshi2015}, we opted for a 5-point scale to ensure items' simplicity and  intelligibility \cite{Likert, Biasutti2017}. 

To offer a common ground to all respondents we introduced the following definition of Artificial Intelligence at the beginning of the questionnaire: ``Artificial intelligence (AI) refers to computer systems that can perform tasks that usually require intelligence (e.g. making decisions, achieving goals, planning, learning, reasoning, etc.). AI systems can perform these tasks based on objectives set by humans with a few explicit instructions.'' Given the heterogeneity of the consulted population, we chose a simple definition that could be intelligible by a large audience.

\begin{table*}[t]
\begin{adjustwidth}{-0.8in}{0in} 
 \centering
 \caption{{\bf PAICE questionnaire design and structure}}
 \begin{tabular}{ | p{0.02\linewidth} | p{0.2\linewidth} | p{0.6\linewidth} | }
 \hline
 & \multicolumn{1}{c|}{\textbf{Question Type}} & \multicolumn{1}{c|}{\textbf{Description}} \\
 \hline
 \parbox[b]{2mm}{\multirow{3}{*}[-7ex]{\rotatebox[origin=c]{90}{\textbf{Awareness}\enspace}}} & Likert scale & \vtop{\hbox{Q1: Knowledge about AI} \hbox{Q3: Impact of AI on daily life (repeated for control question)} \hbox{Q5: Awareness of interaction with products incorporating AI} \hbox{Q7: Awareness of the application of AI in different sectors across Europe}}\\
 \cline{2-3}
 & Dichotomous & \vtop{Q4: Knowledge about three specific European initiatives: the General Data Protection Regulation (GDPR), the Ethics Guidelines for Trustworthy AI, the proposal of an AI regulation}\\
  \cline{2-3}
 & Multi-response & \vtop{Q6: Awareness of products embedding AI}\\
 \hline
 \parbox[t]{2mm}{\multirow{1}{*}[-1ex]{\rotatebox[origin=c]{90}{\textbf{Attitude}\enspace}}} & Likert scale & \vtop{\hbox{Q2: General attitude towards AI} \hbox{Q8: Attitude towards the application of AI in specific sectors} \hbox{Q9: Perceived comfort with a scenario applying AI to job recruitment} \hbox{Q10: Perceived comfort with a scenario applying AI to energy consumption}}\\
 \hline
 \parbox[b]{2mm}{\multirow{2}{*}[-4ex]{\rotatebox[origin=c]{90}{\textbf{Trust}\enspace}}} & Likert scale & \vtop{\hbox{Q12: Importance of specific policy measures to increase trust} \hbox{Q13: Importance of education to increase trust in AI} \hbox{Q14: Trust in entities that may ensure a beneficial use of AI}}\\
 \cline{2-3}
 & Ranking & \vtop{Q11: The three most important ethical requirements derived from \cite{Trustworthy_EU} in relation to the aforementioned scenarios (i.e. Q9 and Q10)}\\
 \hline
\end{tabular}
\label{tab:paice}
\end{adjustwidth}
\end{table*}

\subsection{Statistics}%veronica / teresa

To explore the theoretical dimensions structuring the PAICE (awareness, attitude, and trust) an Exploratory Factor Analysis (EFA) and confirmatory factor analysis (CFA) were performed. The aim was to evaluate the robustness of items in the questionnaire. To do this we randomly split the sample (n=4,006) into two groups n=2,450 for EFA and n=1,051 for CFA.
Note that only items measured on the Likert scale were included in this analysis. 

%The aim  was to evaluate the robustness of items in the questionnaire where only items measured on the Likert scale were included in this analysis (39 items). 
%Therefore,
%To do this we randomly split the sample (n=4006) into two groups n=2450 for EFA and n=1051 for CFA. 
%To assess  the European Perception on AI -EPAI construct validity, an 
The EFA was performed to determine the number of fundamental (latent) constructs underlying the set of items and quantify the extent to which each item is associated with the construct \cite{Tinsley2020}. 
%obtaining information about the underlying constructs \cite{Tinsley2020}. 
In this context, the EFA allows us to study the strength of relations between the dimensions identified by the team of experts who designed the questionnaire and the associated items. 
%and gives us information about alternative structures of the questionnaire based on a different number of latent constructs. 
%This information will be particularly useful to inform further research on the validity of the questionnaire. 
%Note that this analysis, as well reveling the plausibility of the hypothesized (tripartite) structure, can serve to inform further research on the validity of the questionnaire. 
%By performing EFA,  items that were substantially non-normal (absolute univariate skew > 2 and/or absolute univariate kurtosis > 7) [ref] were excluded before to conducting the analysis as well as items that did not correlate at a magnitude of at least .3 with one or more other items were also excluded [ref]. Additionally, two criteria can be used to determine whether factor analysis was appropriate:  the Kaiser-Meyer-Olkin (KMO) index and Bartlett’s test of sphericity have been performed. A KMO index > 0.7 and a Barlett’s test of sphericity p-value < 0.05 176 are considered appropriate values to conduct the EFA. Since the study used an ordinal measure of a four-point Likert scale response in the questionnaire, we assessed the polychoric correlation matrix (REF). 
Before performing the EFA analysis, two criteria were tested to determine whether factor analysis was appropriate: the Kaiser-Meyer-Olkin (KMO) measure of sampling adequacy and Barlett's test of sphericity assumptions \cite{Kaiser1974, Fabrigar1999}. 
A KMO index $>0.7$ and a Barlett's test of sphericity p-value  $<0.05$ are considered appropriate values to conduct the EFA \cite{Kaiser1974,Fabrigar1999}. The implementation of the EFA was based on a polychoric correlation matrix since the questionnaire is composed of ordinal items (i.e. Likert scale). The EFA was run by using principal axis factoring, because it does not assume normality of data \cite{Brown2015}, with oblique rotation. The parallel analysis was used to identify the optimum number of factors to be retained. We also assessed inter-factors correlation, in order to evaluate if  some theoretical dimensions are correlated strongly with each other, i.e. $>0.7$.

%We also  considered the factors based on only factors loading but also the interpretability of the factors. We also assess the inter factor correlation to increase discriminant validity and removed the items with low factor loading ($<0.5$), communalities ($< 0.4$) and cross-loadings ($< 0.15$) at each step of the iterations. \textcolor{blue}{I added the parentheses around the inequalities, e.g., ($<0.5$), hope it makes sense. Are these specific numbers typical? If so we could make it explicit and link to a reference, if not some brief justification may be in order.} 
%Note that we applied  all these criteria since  EFA analysis is an interactive, multi-step process. 
To assess the internal consistency of the EFA solution, we calculated Cronbach’s $\alpha$ and ordinal $\alpha$ which is considered the most appropriate coefficient for ordinal-type scales \cite{Gadermann2012, Zumbo2007}. These indices take values in the range [0, 1], so the internal consistency is acceptable if the indices are greater than 0.80 \cite{nunnally1994}. 

Finally, the validity of the factor structure derived from the EFA was evaluated by using the CFA. The implementation for the CFA was based on a polychoric matrix and the robust diagonally weighted least squares (RDWLS) extraction method which is more suitable for ordinal data than other extraction methods \cite{ch2016}. 
 
We assessed the fit of the model using the following criteria: root mean squared error of approximation (RMSEA$<0.08$), the comparative fit index (CFI) and  the Tucker-Lewis index (TLI) with  values above 0.95 and 0.90, respectively, and the standardized root mean residual (SRMR$ <0.08$) \cite{Taasoobshirazi2016}. 
%In addition, the Chi-squared test was used although this test is sensitive to sample size. 
%but in this case it is valid, as the sample size include n= 1501 people.\\
%All analyses were performed using SPSS software version V.25.0.2 and R version 3.4.1 (The R Foundation for Statistical Computing, Vienna, Austria).
All analyses were performed by using the statistical package for social science V.25.0.2 (SPSS) and R version 3.4.1 (The R Foundation for Statistical Computing, Vienna, Austria).

\subsection*{Limitations}
This work, like any other, has some inherent  limitations. Though we tried to represent different European regions, the sample does not cover all European countries. Thus, our analysis may not be representative of the opinions of all EU citizens. As we suggest in the conclusion, extending the questionnaire to other countries will give a more complete picture of European society. Moreover, our questionnaire administration methodology (CAWI) assumes that the target population has access to the internet and is familiar with web navigation. This choice could have indirectly impacted on the selection of the population interviewed. The latter may be skewed towards people with higher education levels and/or wealthier socio-economic status.  Another limitation concerns the measurement of awareness. In this study, we focus on self-reported awareness, which may suffer from subjective and contextual factors. Objective knowledge about AI is another important dimension of awareness; its rigorous measurement would require the development of a specific methodology going beyond the scope of this work.

\section{Results} \label{results}%atia and cristian

The responses to the questionnaire are presented with respect to the three dimensions: Awareness, attitude, and trust. Aggregated responses to all items are reported in tables ``Likert-scale items'' and ``Non-Likert scales'' with the descriptive statistics in the supplemental material (Responses).
Responses were compared with respect to different groups by using Kruskal-Wallis test where a p-value $< 0.01$ is considered statistically significant. In our comparison, we considered the following groupings: Countries, age groups, and gender. With respect to countries, we found statistically significant differences among groups for all Likert scale items. As for age and gender groups, we found statistically significant differences for a subset of items. For the sake of brevity, in the subsequent sections, we will comment only most significant differences. The results of statistical tests are reported in tables ``Awareness'', ``Attitude'', and ``Trust'' in the supplemental material (Statistics by Groups).

Finally, we report the results of the analysis performed to assess the questionnaire's validity and reliability. 

\subsection{Awareness}

\begin{figure*}[!ht]
\centering
%\begin{adjustwidth}{-1.70in}{0in}
%[width=18cm]
%[width=\linewidth]
\includegraphics[width=\linewidth]{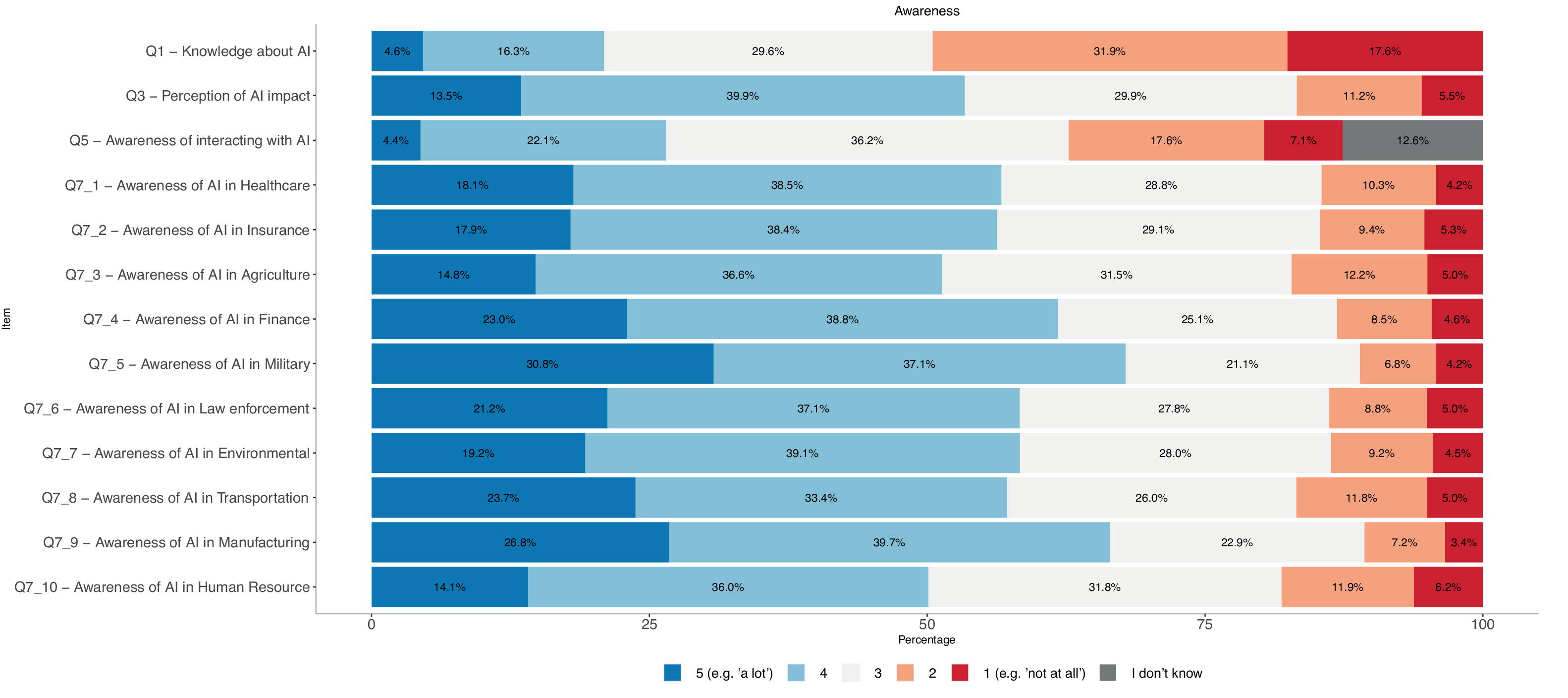}
%\end{adjustwidth}
\caption{Responses to Likert scale items associated with awareness. Low-scale values (1 and 2) are represented by red-like colors, while high-scale values (4 and 5) are represented by blue-like colors. Item Q7 is split into sub-items regarding the perceived presence of AI in ten different sectors. %\new{[Alessandro: If possible, I don't know in Q5 should be the rightmost item (rather than left-most) as it is closer to unawareness than awareness]}
%Item Q7 presents different sectors of application in the following order: 1 Healthcare, 2 Insurance, 3 Agriculture, 4 Finance, 5 Military, 6 Law Enforcement, 7 Environment, 8 Transportation, 9 Manufacturing, 10 Human Resources.
}

\label{Fig_1}

\end{figure*}
%When it comes to AI and its impact on society 49.5\% of the respondents feel like they have almost no knowledge or basic knowledge of competency towards AI. 29.6\% declare having intermediate knowledge while only 20.9\% feel like they have advanced or expert knowledge.

In Figure \ref{Fig_1}, we represented the percentages of responses to Likert scale items connected to awareness. Blue and red colored segments identify the two extreme positions: high and low levels of awareness respectively. 
The largest red segment, including the lowest scale values (i.e. 1 and 2), regards the self-assessed competency on AI (Q1). In this item, almost half of the respondents (49.5\%) reported having low or no knowledge, while only 20.9\% considered their knowledge to be advanced or expert-level.%When asked about being aware of interacting with a product or service based on or including AI (Q5), only 30.3\% are often or always aware, while 28.2\% declared to be never or seldom awareness. %of interacting with an AI-infused product or service, 41.4\% of them says they are aware sometimes, while 30.3\% declare being aware often or always. 
%Regarding the extend that people feel affected by AI and its applications impact in their your daily life
Analyzing the results by country, Germany and the Netherlands have the highest percentage of respondents who feel less knowledgeable, at 66\% and 63\% respectively. If we look at gender, the percentage of individuals who feel less competent is greater for males (55\%) as compared to females (43\%). With respect to age, the portion of individuals with low or no competency is higher for seniors (63\%) and lower for young respondents (32\%).

When asked about being aware of interacting with a product or service based on AI (Q5), only 26.5\% reported being often or always aware, while 24.7\% reported to be never or seldom aware, and 12.6\% chose the ``I don't know'' option. In Germany the fraction of people who feel never or seldom aware increases by 9 percentage points (32\%). Male respondents declared a higher rate of low or no awareness of interaction (25\%) as compared to females (23\%). The group of senior respondents achieved the highest percentage of answers expressing unawareness during interaction (28\%). 
%\new{[Alessandro: this is a general pattern: whenever we group 5-4 and 1-2 from the Likert scale, we can write so a couple times for the reader, but then we can take it for granted. Otherwise we are jumping through hoops to reiterate somehting that is already clear at the expense of readability.]}

In relation to the impact of AI in their daily lives (Q3), half of the respondents (53.4\%) felt like it has somewhat or a lot of impact,%in their daily lives, 29.9\% are neutral and
while 16.7\% answered with ``not so much'' or ``not at all''. The perception of (high) impact is greater in Spain (73\%) and lower in Poland (33\%) - the latter is also the country in which there is the highest fraction of answers reporting a low perceived impact of AI on their lives (29\%).
%We then asked if our participant ever have heard of any European initiatives regarding AI. 

%Regarding their familiarity with the normative and ethical European framework (Q4), two out of three respondents (65.6\%) have heard about GDPR, while only one out of three were aware of the Trustworthy AI Guidelines or the AI Act (28.3\% and 29.8\% respectively) %have heard of the Ethics Guidelines or 29,8\% about the AI Act regulation proposal.
%We also presented a list of applications to understand which one was believed to contained AI according to our respondents.

%We then asked to what extend they think that AI was used in different sectors in Europe.

Items from Q7\_1 to Q7\_10 assessed to what extent respondents feel AI is used in distinct sectors across Europe.
%provides a set of sectors and industry domains aiming to perceive to what extend AI was used across Europe. % provide a list of sectors and industry domains and asked to what extend AI was used in each of them in Europe.
%When asked about the extent to which AI is used in different sectors, 
Military (67.9\%) and Manufacturing (66.5\%) present a higher fraction of respondents perceiving AI as being somewhat or very present in such sectors. On the other hand, Human Resources (50.1\%) and Agriculture (51.4\%) present a lower perception of the presence of AI. 

Regarding respondents' familiarity with the normative and ethical European framework (Q4), two out of three respondents (65.6\%) have heard about GDPR, while only one out of three were aware of the Trustworthy AI Guidelines or the AI Act (28.3\% and 29.8\% respectively).

Participants were also introduced to a list of applications and were asked about which ones may contain AI components (Q6). Facial recognition apps, content and product recommendations, search engines, traffic navigation apps, and car ride-sharing apps were the most identified applications, selected by half of the respondents. Other options with more limited AI applications, such as calculators or text editors, were included by 32.6\% and 26.3\% of participants respectively. Finally, 7.2\% of respondents selected the option ``none of the above'', hence did not identify any AI-based application.

\subsection{Attitude}

\begin{figure*}[!ht]
\centering

\includegraphics[width=\linewidth]{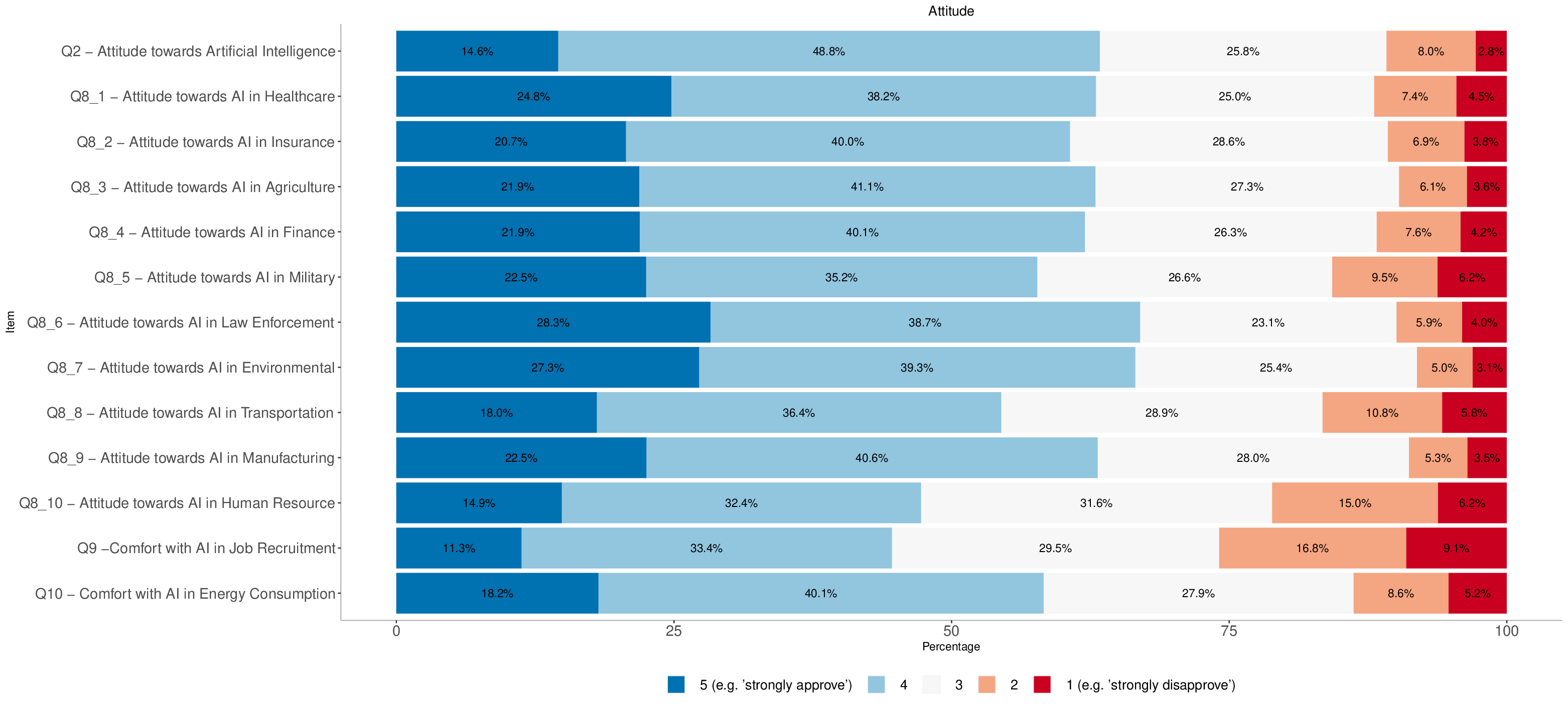}

\caption{Responses to Likert scale items associated with attitude. Low-scale values (1 and 2) are represented by red colors, while high-scale values (4 and 5) are represented by blue colors. Item Q8 is split in sub-items regarding the attitude towards AI in ten different sectors.
%Item Q8 presents different sectors of application in the following order: 1 Healthcare, 2 Insurance, 3 Agriculture, 4 Finance, 5 Military, 6 Law Enforcement, 7 Environment, 8 Transportation, 9 Manufacturing, 10 Human Resources.
}
\label{Fig_2}
\end{figure*}

%We asked our respondent to describe your attitude towards Artificial Intelligence (AI) and its applications?
%In this section, participants were asked to describe their attitude towards AI and its applications. 
In Figure \ref{Fig_2}, we reported the percentages of responses to Likert scale items associated with attitude, where blue segments represent a (very) positive inclination and red segments indicate a (very) negative one.  
%Bars' distribution gives a quick glimpse of respondents' attitude: the closer to the right side the more positive the approach and vice versa.

Regarding their general attitude towards AI (Q2), 63.4\% of the respondents report strongly approving or approving of AI.
%25.8\% were indifferent and 10.8\% disapproved or strongly disapproved its use. 
The most receptive countries were Romania and Spain with almost 80\% approval, while in France fewer than 50\% participants declared approval of AI. With respect to gender, females expressed to be more positive as compared to males, with approval or strong approval at 68\% and 59\% respectively. When considering age, the class of younger respondents reached the highest rate of approval (70\%), while the group of seniors reported the lowest one (58\%).
%We then investigate more in detail in which which were their attitude according to the sector of application in Europe. 

Items from Q8\_1 to Q8\_10 aimed to further understand how approval varies depending on the sector of application.
%how these opinions were distributed in relation to the use of AI in different sectors and industry domains and in to two specific use cases. 
%When asked about the attitude towards the use of AI in different sectors, 
Law Enforcement and Environment have the highest acceptance with an average of 67\% of participants opting for approval or strong approval, followed by Manufacturing, Healthcare, and Agriculture. Human resources presents the lowest acceptance rate (47.3\%) and the highest dissatisfaction rate with 21.2\% of respondents disapproving or strongly disapproving. 
%Moreover, in addition to the question we present two scenario. 

We also considered two specific use case scenarios: Q9 presents an AI-based system that screens candidates' resumes and selects those who can access the interviewing stage \cite{bogen2018help}; Q10 introduces a smart meter to reduce energy consumption inspired by \emph{demand side management} \cite{khan2019smart} that leverages AI to recommend more efficient usage and provide personalized offers from energy providers. 
%related to the human resources and environmental sectors respectively. The former presents an AI-based system that scans candidates' resumes and analyses a set of answers to a job questionnaire assigning a score to their candidacy that is used to select which ones can access to the job interview.
While the proportion of neutral positions is approximately the same for both scenarios, the approval is significantly higher for the smart meter with 58.3\% of the respondents feeling fairly or very comfortable, as opposed to 44.7\% for the resume screening system. Again, we observed statically significant differences among countries. Poland is the most receptive country with about 67\% of respondents feeling fairly or very comfortable in both scenarios. The trend for gender and age groups is similar to that found for general attitude with a preference for the smart meter scenario.

%One out of four participants (25.9\%) considered they feel not at all comfortable or not very comfortable, 29.5\% would feel neutral, while 44.7\% would feel fairly or very comfortable.
%The latter scenario, introduces a smart meter to reduce energy consumption at home empowered with AI that recommends more efficient usage and provides personalised offers from energy providers that can help to save money. 
%In this case, over half of participants (58.3\%) declared to feel very or fairly comfortable, 27.9\% would be neutral while only 13.8\% would feel not very or not at all comfortable.

\subsection{Trust}

\begin{figure*}[!ht]
\centering
\includegraphics[width=\linewidth]{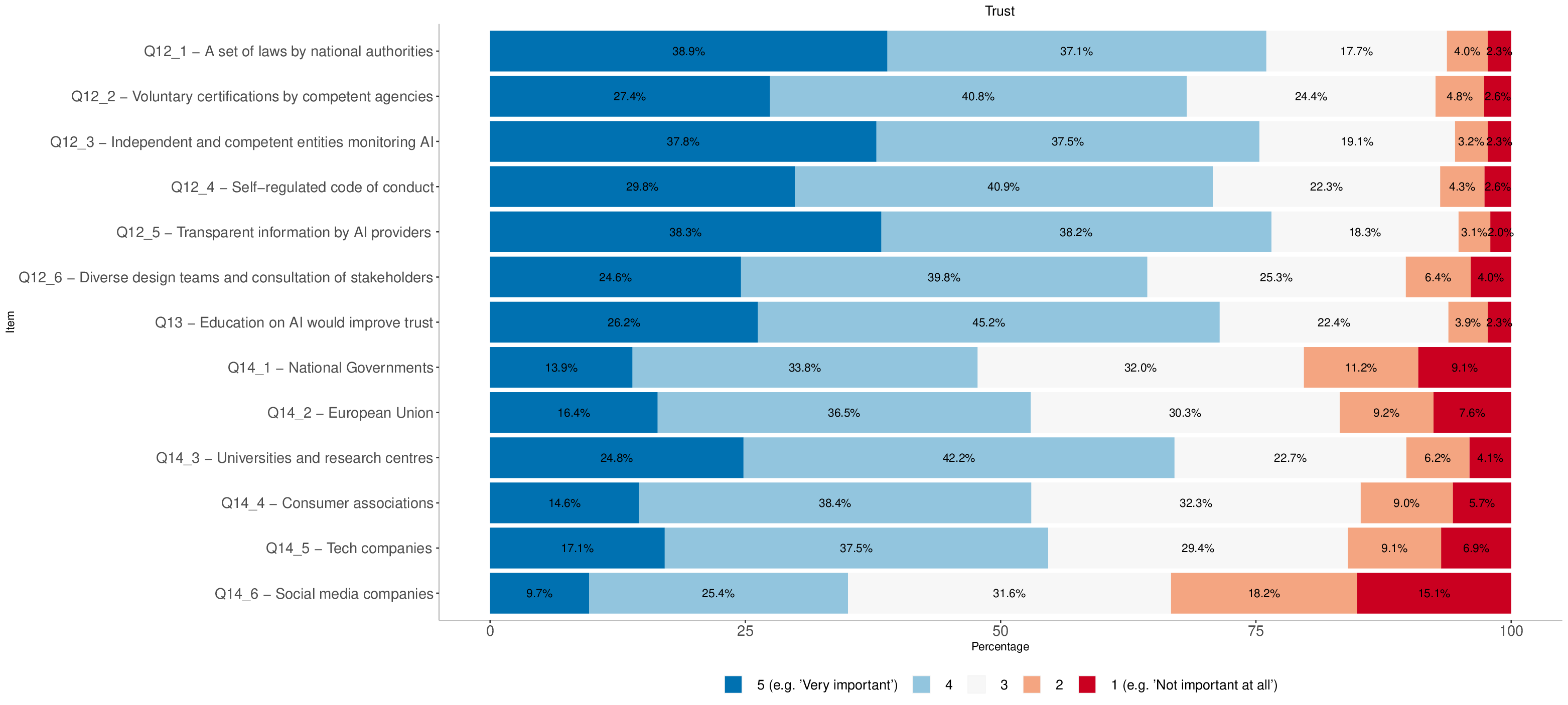}
\caption{Responses to Likert scale items associated to trust. Low-scale values (1 and 2) are represented by red-like colors, while high-scale values (4 and 5) are represented by blue-like colors. Item Q12 is split into sub-items regarding the perceived importance of six different policy measures. Item Q14 is split in sub-items related to the perceived trust in six different entities. 
%Item Q12 refers to the following measures: 1 set of laws, 2 voluntary certification, 3 external audits, 4 code of conduct, 5 transparent information, 6 presents different sectors of application in the following order: 1 Healthcare, 2 Insurance, 3 Agriculture, 4 Finance, 5 Military, 6 Law Enforcement, 7 Environment, 8 Transportation, 9 Manufacturing, 10 Human Resources.
}
\label{Fig_3}
\end{figure*}

In Figure \ref{Fig_3}, we represented the responses to Likert scale items referred to trust. Similarly to previous dimensions, colors are indicative of respondents' satisfaction with actions and entities aimed to ensure trust.
When asked to assess the importance of a set of policy measures to increase trust (Q12), 76\% of the respondents valued as important or very important the deployment of a set of laws by a national authority that guarantees ethical standards and social responsibility in the AI application. Romania and Germany are the countries in which this percentage is the largest, at 90\% and 82\% respectively. With regard to age, a large proportion of senior respondents consider this measure important (81\%), followed by young (71\%) and the middle age respondents (68\%).
%three out of four participants valued very important or important the deployment of a set of laws by a national authority that guarantees ethical standards and social responsibility in the AI application (76\%), the provision of clear and transparent information by the provider that describes the purpose, limitations and data usage of the AI product (76.6\%), and having independent expert entities that monitor the use and misuse of AI in society, including the public sector, and inform citizens (75.4\%). 
The remaining measures were also highly supported (more than 50\%); the least valued one was the creation of diverse design teams and the consultation of different stakeholders throughout the entire life cycle of the AI product (64.4\%). Education as a remedy to improve citizens' trust (Q13) was also largely approved with 71.4\% of agreement or strong agreement. Note that this percentage increases significantly in Romania and Spain where agreement reaches 85\% and 83\% respectively, while it falls to 59\% in France. 

With respect to trusted entities ensuring a beneficial use of AI (Q14), two out of three participants (67\%) rated universities and research centers as entities that could be trusted a lot or somewhat. Note that this percentage varies across countries with Romania reporting the highest value (77\%) and France the lowest one (55\%). 
Social media companies are the least trusted entity with only 35\% of respondents trusting them. With respect to countries the percentage is higher in Italy (47\%) and lower in the Netherlands (26\%), while, if we consider age groups, trust in social media is lowest for senior respondents (24\%) and highest for young respondents (46\%). 

%which obtained similar results as trusted entity (33.3\%) and not trusted so much or at all (35.1\%). 
With Q11, we asked to select three out of the seven most important aspects that an organization should consider to developing or using AI in relation to the previous scenarios (Q9 and Q10).
%based on the principles of the Ethics guidelines for Trustworthy AI(Q11). %The respondents reported as first choice with 30.8\% Privacy and Data Protection, 20.4\% voted for Security and Accurate results, and 15.6\% opted for Human supervision over the outcome process. As second choice, 21.7\% chose Privacy and Data Protection, 17.3\% Human supervision over the outcome process, and 16.9\% Security and Accurate results. As third choice, 17.3\% Clear communication about the AI application, purpose and limitations, 16.7\% Human supervision over the outcome process, and 15.4\% Security and Accurate results. 
Interestingly, there is a clear preference towards technical aspects related to security, robustness and human oversight, with Privacy and Data Protection leading as a choice for 30.8\% respondents. On the other hand, the Societal and Environmental impact of AI applications was only selected by the 5\% of the respondents as a first or second choice.

\subsection{Questionnaire validity and reliability} %veronica e teresa

%The data analysed are based on an ordinal scale where participants responded to an item by indicating a level of measurement, defined on 5-point Likert scale.
Among the 4,006 participants, 501 ticked the response option ``I don't know'' for item Q5 (i.e. ``How often are you aware of interacting with a product/service based on or including AI?''), corresponding to 12.6\% of the sample. Therefore, these responses were excluded from the statistical analysis. A qualitative analysis was conducted to explore the content of each item and identify the ones with multicollinearity issues \cite{cohen2017,joreskog1994}.
%and Items with high and low polychoric intercorrelations were selected. In particular, the correlation % matrix was scanned considering  a quantitative criterion for intercorrelations that is  |r| > 0.80 and |r| < 0.35. \\ 
%Based on exploratory factor analysis (EFA) on a random subsample (N = 2450), we proposed a model that was subsequently tested on the rest of the sample (N = 1051) using confirmatory factor analysis (CFA). 

The Kaiser–Meyer–Olkin (KMO) test and the Bartlett’s test of sphericity showed that the data are appropriate to perform the EFA with a KMO index $=0.93$ and a Barlett's test of sphericity p-value  $<0.0001$ \cite{Kaiser1974,Fabrigar1999}. 
Parallel analysis suggested three factors which are detailed in the supplemental material (Exploratory Factor Analysis: Figure). The three factors accounted for 62\% of the total variance. In particular, we extracted the factors based on factor loading and the interpretability of the factors. Note that, the items with low factor loading ($<0.50$) were not considered while the remaining items were assigned to a single factor according to their highest loading (see supplemental material (Exploratory Factor Analysis: Table).

The items that load on the same factor suggested that factor 1 (26\% of the total variance) refers to awareness and includes 7 items (Q7\_2, Q7\_4, Q7\_5, Q7\_6, Q7\_7, Q7\_9, Q7\_10); factor 2 (25\% of the total variance) refers to attitude and includes 6 items (Q8\_1, Q8\_2, Q8\_3, Q8\_4, Q8\_6, Q8\_7) and factor 3 (10\% of the total variance) refers to trust and includes only 3 items (Q14\_2, Q14\_4, Q14\_6).

We also assessed the factor correlation matrix of the final EFA to assess the discriminant validity. The correlations between all three factors were found positive. The largest positive correlation was between factor 1 and factor 2 (0.52), and the smallest correlation was between factor 2 and factor 3 (0.37).  Hence, we did not find correlation coefficients greater than 0.7; therefore the factors derived from EFA revealed adequate discriminant validity among the factors.

For reliability, both Cronbach’s $\alpha$ and ordinal $\alpha$ were found to be large enough ($\alpha > 0.8$), indicating that the questionnaire had good internal consistency. Then, we used the CFA to examine the proposed factorial structure of the PAICE.  
Overall, our CFA results showed that the EFA model showed  acceptable fit indices (RMSEA (90\%CI = 0.011; CFI = 0.99; TLI = .99, SRMR = 0.03; p-value $<0.001$).

\section{Discussion}
\label{discuss}

The collected responses reveal some contrasts that are worthy of in-depth analysis. These tensions may signal friction in current efforts towards a Trustworthy AI innovation and, in particular, call for reflection on the EU context, where the AI strategy aims to build an ecosystem of trust and the development of an AI regulation is underway.  
Note that these contrasts reflect more implicit contradictions rather than disagreements openly expressed. 
Yet, pointing them out allows us to discuss critical social orientations that may constitute a barrier to the development of a trustworthy AI culture and, most importantly, an inclusive approach to AI governance.

\subsection{Implicit contradictions}

\paragraph{\textbf{Knowledge about AI \emph{vs} Approval of AI}.}
The first remarkable result of this survey is that respondents' (self-assessed) knowledge of AI is much lower as compared to their approval, which is, by contrast, quite high with respect to both AI generally considered and several domain applications. This confirms other studies that observed a gap between people's limited competency and their perceptions and expectations, which might be influenced by the narratives about the future progress of emerging technologies such as AI \cite{TRS}.

\paragraph{\textbf{AI for the environment \emph{vs} The environmental impact of AI}.}
AI approval is often dependent on the sector or context of application, such as education or healthcare. In this respect, the high acceptance rate of AI in law enforcement and the environment is rather striking. A plausible interpretation might be that people consider these as critical areas where the use of advanced technologies, like AI, could ensure greater progress as compared to other sectors. However, it is surprising that only a small portion of the respondents choose societal and environmental aspects as one of their ethical priorities. In other words, it seems that the intuition of the beneficial effect of AI on important environmental challenges ahead is not on par with the knowledge of possible negative impacts that AI may have on society and the environment. This intuition would be in line with previous studies showing that people tend to not care about the environmental impact of AI solutions and pay more attention to transparency and explainability \cite{konig2022consumers}.

\paragraph{\textbf{Perceived AI impact \emph{vs} Knowledge about EU measures on Trustworthy AI}.}
While the perceived impact of AI is high across the interviewed population, the knowledge of recent measures put forward by the EC to safeguard the risks associated with the use of AI is significantly low. In particular, about 70\% of the respondents claim no knowledge about two key recent actions by the EC, i.e. the ethics guidelines for Trustworthy AI \cite{Trustworthy_EU} and the proposal for an AI regulation \cite{ai_act}, whereas most of them are familiar with the GDPR. Though this lack of knowledge can be partially explained by the novelty of these initiatives (April 2019 and April 2021 respectively), it seems that the public discussion of the AI impact in the EU is still remote from citizens' experience. Also, the lack of knowledge about the proposal for an EU regulation on AI is somewhat in contrast with respondents' policy preferences indicating the introduction of laws as a top priority. 

\paragraph{\textbf{Introduction of laws by national authorities \emph{vs} Trust in national governments}.}
As anticipated, the set up of laws by national authorities is acknowledged as (very) important by the largest portion of the respondents. However, national governments are the second last entity that can be trusted a lot or somewhat. This last opinion may reflect a larger discontent with democratic processes \cite{Cordis}, challenged by global crises (e.g. climate, migration, economy etc) and more recently by the Covid-19 pandemic. The EU took a leading position in proposing global standards for the governance of AI and promoting a unified approach to AI across all member states. However, the implementation of these policy and regulatory efforts might be undermined by the fragile relations between citizens and democratic institutions and associated phenomena (e.g. anti-EU sentiments and populist movements).

\paragraph{\textbf{AI Education as measure to improve citizens' trust \emph{vs} Interest in engaging with AI education}.}
With respect to the role of education in fostering trust in AI, 71\% respondents are highly positive and express a (strong) approval. The value of education and culture is also reflected by the choice of universities and research centers as the most trusted entities in ensuring the beneficial development of AI. To gain a better understating of the value of education we also asked participants if they would be interested in attending a free course on AI with a view to improve their knowledge (see the last question, Q16). Overall, 61\% of participants answered positively, although compared to their strong support of education-related initiatives, even higher percentages could be expected. Moreover, only half of those who self-reported a low AI competence (Q1) said they would be interested in attending a free course (Q16). This limited interest in engaging with AI education might be indicative of a sort of hesitancy in joining the innovation process brought about by AI, in particular among individuals who feel less competent. A similar interpretation may also apply to the selection of inclusive design teams and consultation with stakeholders (Q12\_6) as the least valued measures.

\subsection{Potential barriers and recommendations}

The combination of the implicit contradictions presented above suggests three interrelated social trends. These may affect the way in which AI innovations integrate into the fabric of social life and create a barrier to the human-centric approach that the EU wishes to achieve. For each trend, we discuss critical issues that policy makers could face and suggest a few recommendations. We recall that the European AI strategy pursues three fundamental goals: 1) boosting the AI uptake across the economy by private and public sectors; 2) preparing for socio-economic changes brought by AI transformations; 3) ensuring appropriate ethical and legal framework to promote trustworthy AI \cite{ai_agenda}.

\subsubsection{\textbf{Approval of a hyped, but poorly known, tech}.}
The divide between knowledge and approval, regardless of its causes, calls for reflection on the meaning and implications of approving something which is not sufficiently known or understood. Over the last few years we have witnessed an explosion of fictional and non-fictional AI-related communication and narratives. This large availability of information sources can contribute to creating big expectations, on the one hand, but can also increase confusion or even resistance and aversion \cite{longoni2019resistance, dietvorst2015algorithm}, on the other hand. For example, \cite{10.5555/3298239.3298381} analyzed trends in beliefs, interests, and sentiment articles around AI in a 30-years period. Results show a significant increase in content with a generally optimistic perspective since 2009, although certain topics regarding ethical, technical, and social aspects of AI are also gaining relevance. 
Moreover, the language used to communicate is highly influential \cite{TRS}; when mixed with fictional, or utopian narratives, it can create confusion and lead the general public to overestimate the real capabilities and limitations of AI, augmenting the disconnect from the real progress of the technology. 

A manifest example of the risk of this poorly informed approval is the attraction created by the latest language model ChatGPT. It seems plausible that, in the imagination of a non-specialist, an AI system of this kind, which creates poems, codes, and answers complex questions in a credible manner, is likely to be credited with advanced cognitive abilities. The problem is that systems like ChatGPT ``can fool us into thinking that they understand more than they do'' \cite{chatGPT}, and this limitation is probably unknown by the majority of users. 
The language and terminology used are fundamental to avoid inaccurate and biased messages that create overhype and misinformation about the real capabilities, limitations and associated risks of AI. Moreover, information needs to be clear and adapted to the audience. 
%Improving contents in media can help gaining general interest and awareness in AI and enable informed debates around progress of the technology, and adoption of the forthcoming laws and policies. 
To improve media communication on AI and support more informed opinion we recommend: 1) increasing the study of media communication on AI and social dynamics created by AI-related content; 2) fostering training of science and tech journalists/communicators on AI applications, in particular, on new AI breakthrough; 3) distributing high-quality information through institutional channels (e.g. curating the terminology and translating material in national languages).

\subsubsection{\textbf{Disconnect from public AI policies}.}
 In democratic societies, institutions play a crucial role in anticipating risks and taking preventive actions to protect citizens' rights when innovation processes take place. This is particularly important in times of global crisis or rapid changes and when parts of the population lack the expertise to face complex challenges \cite{zwitter2012rule}. However, the development and implementation of public policies and laws are more effective when citizens participate in the public discussion and gain a better understanding of the issues at stake \cite{funk2009public}. Indeed, increasing public awareness may impact people's values and priorities. Not surprisingly, privacy and data protection, which turned out to be the most well-known EU action (i.e. the GDPR), is one of the highest-rated ethical requirements by the respondents of the survey. We recall that before the GDPR was released there were already a directive and respective national laws regarding data privacy. Moreover, the regulation was accompanied by a large campaign of information and awareness towards the topic, in addition to a two-year adoption period for companies (from 2016 to 2018). 
 
 The implicit contrasts observed in our results stress the need of supporting European citizens in gaining a greater understating of the risks associated with AI, including harms that might be invisible to them. In particular, more efforts should be made to raise awareness of AI’s environmental costs in the public discourse as suggested by \cite{brevini2020black}. A poor understanding of societal harms associated with AI may contribute to exacerbating inequalities and eroding democratic processes. Moreover, if people have limited knowledge about the rules and the initiatives introduced to protect them from potential AI-related risks, they will not be aware of the rights they have and when these are violated. Overall, reflecting on the gap between citizens and the EU policy efforts on AI stresses the importance of building a culture of trust on top of laws and policies. Educational and dissemination resources are needed to promote the last key EU policy initiatives and to empower citizens to know their rights and exercise them. Greater attention should also be directed to the initiatives of inclusive governance to avoid the so-called paradox of participation \cite{cleaver1999paradoxes, png2022tensions}, i.e. inclusion processes failing to achieve structural reforms.
 To improve participation and make society a relevant stakeholder in AI policy-making we recommend 1) analyzing the effectiveness of EU initiatives and platforms aimed at  stakeholders' participation, including the European AI alliance \cite{AIAlliance}; 2) creating information material on the AI-related risks and associated EU measures targeted to different audiences (e.g. children and seniors); 3) increasing local initiatives (including physical events) on AI and the EU efforts aimed at reaching segments of society who are at the edge of current AI debates (e.g because lacking technology or other cultural resources).

 \subsubsection{\textbf{Poor engagement with AI education and training}.}
Education and lifelong learning play a central role in the European AI policy. These strategies aim at boosting economic growth but also preparing the society as a whole, to ensure that ``none is left behind in the digital transformation'' \cite{ai_agenda}. This preparation includes the introduction of AI from the early stages of education to increase skilled workers in AI-related tasks, but also the promotion of conditions that make Europe able to attract and retain talent for AI research and industry. These steps connect with the need to preserve democracy and core values in our society increasingly shaped by AI, big data, and behavioral economics \cite{helbing2019will, cristianini2020social}. 
As well as the Digital Europe Programme (DIGITAL), the EU supports several projects to train AI experts and stimulate excellence (e.g. TAILOR, ELISE, HumanE-AI-Net). European countries are also making efforts to improve AI education at a national level as reported in their AI strategies \cite{Foffano22} and the AI Watch investment dashboard - apparently the investments made in talent, skills, and lifelong learning represent about 60\% of total investments by private and public organizations \cite{AIwatch}.

%(see more details on the AI Watch Dashboard \url{\https://ai-watch.ec.europa.eu/ai-investments-dashboard_en}).
While it is widely acknowledged that education and training are key in promoting citizens' participation, what such an education should look like is open to discussion. This problem regards the type of knowledge and skills that we will value in the future. As the economy increasingly relies on AI, we expect that AI-related skills, such as algorithmic formalism  \cite{green2020algorithmic}, will take a greater role in education and culture. However, this change may favor critical processes such as the prioritization of algorithmic thinking over other forms of knowledge \cite{boyd2012critical} and the subordination of education to business and economic interests. Note that the influence of economic drivers in the shaping of AI education could also damage the very field of AI by increasing the role of techniques and approaches with a higher economic and commercial impact and marginalizing the others.
Another issue regards how to deliver AI education and training. Several resources are available online supporting self-education on AI, many offered for free \cite{elemAI}. However, if this becomes the default option, some people might be excluded from AI education, such as workers who have a low level of formal education and digital skills. We should also consider to what extent people feel comfortable with the education offered, whether they experience anxiety or social pressure. Further concerns regard courses offered by big tech companies and how these can influence the public discourse on AI as well as AI research \cite{luchs2023learning}.

To address the issues connected to the shaping of AI education we recommend 1) assessing to what extent people feel conformable with existing educational resources on AI and identifying categories of the population that might be excluded; 2) increasing the integration of the humanities into computer science and AI curricula to help future tech people address broader socio-technical challenges; 3) reconsidering the incentives of research careers, now dictated by the dynamics and standards of individual disciplines, in light of multidisciplinary collaborations and societal challenges raised by techno-science.

\section{Conclusions}
\label{conclude}
This paper presents and discusses the results obtained from the PAICE questionnaire. 
The collected responses show that European citizens have low knowledge of AI capabilities in different applications and domains, as well as of the efforts aimed to build an ethical and regulatory framework for this technology. The analysis of our results suggests some tensions connected to broader social trends that lead to reflection on aspects that may interfere with policy efforts towards Trustworthy AI: 1) an uninformed approval recalls attention to the risks of misinformation and poor narratives about AI; 2) a disconnect from EU policy on AI brings attention to the need of high-quality communication campaigns on the AI-related harms and current EU policy and regulatory efforts; 3) a poor engagement with AI education and training strategies points to the risks of growing social and cultural inequalities.  

Through the analysis of the validity and reliability of the questionnaire (PAICE) we assess the robustness of the theoretical structure identified by the working group during the design process and support the research community in the reuse of the PAICE. The validation of the questionnaire shows that for a subset of items, PAICE can be used to measure awareness, attitude, and trust towards the AI ecosystem.
In addition, PAICE proves useful in providing respondents with new stimuli making them reflect upon their interaction with new technologies and its possible impact on society. At the end of the questionnaire, we repeated item Q3 investigating the perceived impact of AI, and found that 62.2\% of respondents answered that AI has an impact on their daily life, an increase of 10 percentage points.
In future work, we plan to extend the questionnaire to new regions and investigate country-specific differences with available data on the AI landscape \cite{AIwatch2021}.

\section{Supplementary Material}

In the following subsections, we provide the links to the supplemental material of the present research work. 

\subsection{Questionnaire}
\label{S1_Questionnaire}
Text in full of the questionnaire on the Perceptions of AI by the Citizens of Europe  (PAICE) translated in English: \url{https://github.com/EU-Survey/Material/blob/main/S1_quest.pdf} 

\subsection{Responses}
\label{S2_Tab_Resp}

Table with all aggregated responses to likert-scale / dichotomous / multi-response items and rankings: 
\url{https://github.com/EU-Survey/Material/blob/main/S2_res.xlsx}. 
For likert scale items, some descriptive statistics are also reported.  

\subsection{Digital skills}
\label{S3_Tab_Dem}

Table with aggregated responses related to digital skills, education and population size grouped by countries: \url{https://github.com/EU-Survey/Material/blob/main/S3_dem.xlsx}

\subsection{Statistics by Groups}
\label{S4_Tab_Comp}

Table with all responses to likert scale items aggregated by countries / age groups / gender with p-values: \url{https://github.com/EU-Survey/Material/blob/main/S4_comp.xlsx}
Responses are presented with respect to the dimension considered (awareness, attitude, and trust).

\subsection{Exploratory Factor Analysis: Table}
\label{S5_Tab_EFA}

Table with the results of the Exploratory Factor Analysis: \url{https://github.com/EU-Survey/Material/blob/main/S5_efa.xlsx}. Exploratory factor analysis (EFA) is based on the polychoric matrix which uses principal axis factoring with oblique rotation

\subsection{Exploratory Factor Analysis: Figure}
\label{S6_Fig_EFA}
Plot of Exploratory Factor Analysis: \url{https://github.com/EU-Survey/Material/blob/main/S6_efa.pdf}

\bibliographystyle{plain}
\bibliography{biblio}

\end{document}